\documentclass[twocolumn,amsmath,amssymb,prb]{revtex4}
\usepackage{graphicx}% Include figure files
\usepackage{dcolumn}% Align table columns on decimal point
\usepackage{bm}% bold math
\usepackage{color}

\def\rc#1{{\color{black} {#1}}}

\newcommand \boldeta {\mbox{\boldmath $\eta$}}

\def \be {\begin{equation}}
\def \ee {\end{equation}}
\def \ben {\begin{eqnarray}}
\def \een {\end{eqnarray}}

\begin{document}

\bibliographystyle{../prsty}

\title{Non-uniqueness of quantum transition state theory and general dividing surfaces in the path integral space}

\author{Seogjoo Jang\footnote{Email:sjang@qc.cuny.edu}}
\affiliation{Department of Chemistry and Biochemistry, Queens College, City University of New York, 65-30 Kissena Boulevard, Queens, New York 11367\footnote{mailing address}  \& PhD programs in Chemistry and Physics, and Initiative for the Theoretical Sciences, Graduate Center, City University of New York, 365 Fifth Avenue, New York, NY 10016, USA}
\author{Gregory A. Voth\footnote{Email:gavoth@uchicago.edu}}
\affiliation{Department of Chemistry, James Franck Institute, and Institute for Biophysical Dynamics, The University of Chicago, 5735 S. Ellis Avenue, Chicago, Illinois 60637, USA}

\date{Accepted to {\it the Journal of Chemical Physics} on April 11, 2017}

\begin{abstract}
Despite the fact that quantum mechanical principles do not allow the establishment of an exact quantum analogue of the classical transition state theory (TST), the development of a quantum TST (QTST) with a proper dynamical justification, while recovering the TST in the classical limit, has been a long standing theoretical challenge in chemical physics.  One of the most recent efforts of this kind was put forth by Hele and Althorpe (HA) [ J. Chem. Phys. {\bf 138}, 084108 (2013)], which can be specified for any cyclically invariant dividing surface defined in the space of the imaginary time path integral.  The present work revisits the issue of the non-uniqueness of QTST and provides a detailed theoretical analysis of HA-QTST for a general class of such path integral dividing surfaces.  While we confirm that HA-QTST reproduces the result based on the ring polymer molecular dynamics (RPMD) rate theory for dividing surfaces containing only a quadratic form of low frequency Fourier modes, we find that it produces different results for those containing higher frequency imaginary time paths which accommodate greater quantum fluctuations.  This result confirms the assessment made in our previous work [J. Chem. Phys. {\bf 144}, \rc{084110} (2016)]  that  HA-QTST does not provide a derivation of RPMD-TST in general, and points to a new ambiguity of HA-QTST with respect to its justification for general cyclically invariant dividing surfaces defined in the space of imaginary time path integrals.   Our analysis also offers new insights into similar path integral based QTST approaches.
\end{abstract}

\maketitle

\section{Introduction}
Certain concepts in classical mechanics cannot be carried over to the quantum regime no matter how useful they may be. Transition state theory (TST)\cite{eyring-jcp3,wigner-jcp5} in its rigorous definition\cite{wigner-jcp5,pechukas-r1,pechukas-jcp58,chandler-jcp68,miller-acc26} is tied to such concepts of classical mechanics, namely, the definability of deterministic trajectories in the phase space. Therefore, no exact analogue of the classical TST can be found in the quantum regime.\cite{miller-acc26,mclafferty-cpl27}  Nonetheless, attempts to develop a quantum TST (QTST) have remained an ongoing and challenging theoretical subject.  This is because there is a significant benefit in calculating a rate constant without carrying out any actual real time quantum dynamics calculation.  The insight available from a quantum correction factor for the TST also serves as an important motivation.   However, a major conceptual issue to be settled in developing a QTST is that distinct starting points of the rate formulation,\cite{mclafferty-cpl27,yamamoto-jcp123,miller-jcp79,affleck-prl46,gillan-jp-c20,voth-jcp91,hansen-jcp,stuchebrukhov-jcp95,cao-jcp105,pollak-jcp108} which all lead to the same TST rate expression in the classical limit, can produce different results in the genuine quantum regime. 
In fact, this non-uniqueness of QTST is not unexpected.  It rather reflects the fundamental principles of quantum mechanics, and should be recognized as an intrinsic quantum feature.  In this sense, it is imperative to understand clearly the assumptions and approximations involved in a given QTST because such an understanding offers a  correct assessment of the implications and utility of that particular QTST. 

In a recent work,\cite{jang-jcp144} \rc{ we analyzed a new kind of QTST formulated by Hele and Althorpe (HA).}\cite{hele-jcp138}  Our analysis offered a context for this theory within the general consensus on the QTST, as described above, by clarifying key assumptions implicit in HA-QTST.  First, we pointed out that the starting expression of HA-QTST\cite{hele-jcp138,althorpe-jcp139} was missing a clear physical origin tied to response or scattering theories, unlike those introduced by Yamamoto\cite{yamamoto-jcp123} and by Miller and coworkers.\cite{miller-jcp79}  Second, we analyzed\cite{jang-jcp144}  an apparent  approximation HA employed for the evaluation of their flux-side correlation function,\cite{hele-jcp138} and have also provided an alternative and exact evaluation. The resulting form based on our evaluation suggested that HA-QTST and ring polymer molecular dynamics (RPMD) TST\cite{craig-jcp122,craig-jcp123} are different in general. 

An important issue that was not discussed in detail in our work\cite{jang-jcp144} was the concept of a {\it general dividing surface defined in the imaginary time path integral space}, which was introduced as a new means to optimize RPMD-TST by Richardson and Althorpe\cite{richardson-jcp131} and had also previously been considered by others\cite{cao-jcp105,mills-cpl278} in the context of improving the path integral QTST (PI-QTST).\cite{voth-jcp91}   It is important to first note that the introduction of such a dividing surface is a departure from the original formulation of the RPMD rate theory by Craig and Manolopoulos\cite{craig-jcp122,craig-jcp123} because not every dividing surface in the imaginary time path integral space can be expressed as a ring polymer average of classical ones, namely, an equal weight linear combination of classical dividing surfaces over all the beads of the ring polymer.    For this reason, our previous analysis\cite{jang-jcp144} was limited to the centroid dividing surface, an obvious ring polymer averaged quantity.  For this latter case, we have confirmed that HA-QTST reproduces  the PI-QTST rate expression above the crossover temperature,\cite{voth-jcp91}  to which the RPMD-TST expression proposed a number of years later was also shown to be equivalent.\cite{craig-jcp123}  \rc{We note that this outcome is not an exact derivation of PI-QTST or RPMD-TST because neither of these formulations of QTST are exact except in certain limits.}
 In a more recent article,\cite{hele-jcp144} HA provided further analysis of the general expression obtained in our work,\cite{jang-jcp144} and presented a result that seems to suggest again the equivalence between HA-QTST and RPMD-TST\cite{richardson-jcp131} for general dividing surfaces.   However, as will be shown here, this is not always true. 

Before presenting our new analysis addressing the issue of general path integral dividing surfaces, we want to stress that our assessment\cite{jang-jcp144} of HA-QTST as being one of many non-unique versions of QTST  remains intact.  In particular, our assessment of the assumption implicit in employing the $t=0_+$ limit of a specially constructed flux-side time correlation function, namely, one that neglects a certain part of the Feynman diagram and thus in effect assumes that certain operators can commute, requires no further clarification. Without such an assumption, and without any coarse-graining in time, the exact $t=0_+$ limit of the flux-side correlation function will always vanish in the quantum limit.\cite{costley-cpl83,wolynes-prl47}  With this issue clarified, we will here focus our attention mostly on the subject of general path integral dividing surfaces. These general dividing surfaces arise from the specially constructed time correlation function within the approximation of HA-QTST. 

An important condition underlying the analysis by HA in their recent work\cite{hele-jcp144} is that the dividing surface is a smooth function of imaginary time, an important assumption that \rc{should not be overlooked.}   We will show here that this analysis does not hold true for more general \rc{ and} yet cyclically invariant dividing surfaces in the imaginary time path integral space.  \rc{This result} confirms our previous conclusion that HA-QTST\cite{hele-jcp138} is not an exact derivation of RPMD-TST\cite{craig-jcp122,craig-jcp123,richardson-jcp131} in general because an exact quantum mechanical identity should be independent of the nature of the dividing surface.  

This paper is organized as follows.  Section II provides a brief overview of HA-QTST and our recent analysis, and summarizes its relationship to RPMD-TST.  Section III provides a detailed evaluation of our final expression for HA-QTST for two types of cyclically invariant paths, and demonstrates our central point that HA-QTST does not lead to RPMD-TST for general cyclic paths.  Section IV provides concluding remarks.    

\section{Summary of HA-QTST and its relationship with RPMD-TST} 
The formulation of HA-QTST,\cite{hele-jcp138} for the case where the population function is the step function $\Theta(x)$ \rc{and assuming one dimensional reaction coordinate}, starts from the following generalized Kubo-transformed side-side correlation function (GKSCF):
\ben
&&\tilde C_{ss}(t) = \int d{\bf q}\int d \boldeta\int d{\bf z}\ \rho({\bf q},\boldeta){\mathcal G}({\bf q}, \boldeta,{\bf z};t)\nonumber \\
&&\hspace{1in} \times   \Theta(\left (f({\bf q})-d\right) \Theta \left (f({\bf z})-d\right)   \ ,  \label{eq:c_sst}
\een
where $f({\bf q})$ represents the dividing surface\endnote{\rc{Although $f({\bf q})$ is defined in the space of imaginary time path integral, this may not always be a true dividing surface of the multidimensional ring polymer potential energy profile. Rather, this can be viewed as a some kind of path integral generalization of the classical dividing surface in general.}} defined in the space of imaginary time path integral, and 
\be
\rho({\bf q},\boldeta)=\prod_{k=1}^P\langle q_{_k}-\frac{\eta_{_k}}{2}|e^{-\beta \hat H/P}|q_{_{k+1}}+\frac{\eta_{_{k+1}}}{2} \rangle \ , \label{eq:def-rho}  
\ee
with $\beta=1/(k_BT)$, and
\ben
&&{\mathcal G}({\bf q}, \boldeta,{\bf z};t)=\prod_{k=1}^P \langle q_{_k}+\frac{\eta_k}{2}|e^{it\hat H/\hbar}|z_k\rangle  \nonumber \\
&&\hspace{1.in} \times \langle z_k|e^{-it\hat H/\hbar}|q_{k}-\frac{\eta_k}{2}\rangle \ . \label{eq:def-g}
\een
As mentioned in the Introduction and in our previous work,\cite{jang-jcp144} the quantum dynamical meaning of Eq. (\ref{eq:c_sst}) is not clear as it is not derived from, {\it e.g.}, any response theory.  The primary motivation provided by HA\cite{hele-jcp138} for using this time correlation function is that the $t=0_+$ limit of its time derivative is nonzero and positive, although the value of the exact time correlation function at time $t=0$ is zero.\cite{jang-jcp144}   \rc{The purely} mathematical basis for the classical-like singularity in the correlation function that imparts its nonzero value at $t\rightarrow 0_+$ is easy to understand, \rc{but its physical meaning is much less clear in the context of quantum mechanics.}  HA then utilized this singular behavior of the time derivative of GKSCF to define the following rate expression:\cite{hele-jcp138} 
\ben 
k_{HA}Z_a=\left . -\frac{d}{d t} \tilde C_{ss}(t)\right |_{t=0_+} \ ,\label{eq:dc_sst}
\een
and evaluated this by employing a short time approximation for real time propagators, $e^{\pm it\hat H/\hbar}$, within the definition of ${\mathcal G}({\bf q}, \boldeta, {\bf z};t)$, Eq. (\ref{eq:def-g}).  \rc{In this procedure, a natural classical-like variable defined as} 
\be
p_k=m (z_k-q_k)/t \ , \label{eq:pk_def}
\ee  
\rc{emerges and plays an important role in the derivation of the final rate expression by HA.\cite{hele-jcp138}} 
However, on the other hand, we have shown\cite{jang-jcp144} through partial integration that Eq. (\ref{eq:dc_sst}) can be expressed as 
\ben
&&k_{HA}Z_a= \frac{1}{(2\pi \hbar)^P}\int d{\bf q}\int d \boldeta \int d{\bf p}  \rho({\bf q},\boldeta) \nonumber \\
&&\hspace{.2in}\times\exp (i{\bf p} \cdot \boldeta /\hbar) \Theta \left ({\bf p}\cdot \nabla f({\bf q}) \right) \delta (f ({\bf q})-d) \nonumber \\
&&\hspace{.2in}\times \sum_{k=1}^P \frac{\partial f({\bf q})}{\partial q_k} \frac{1}{2m}\left (\tilde p_{k,+}+\tilde p_{k,-} \right) \ , \label{eq:dc_sst-10+2}
\een
where $\tilde p_{k,+}$ and $\tilde p_{k,-}$ are average imaginary time momenta defined as follows:
\ben
&&\bar p_{k,+}= \frac{\langle q_{k-1}-\eta_{k-1}/2|e^{-\beta\hat H/P}\hat p|q_k+\eta_k/2\rangle}{\langle q_{k-1}-\eta_{k-1}/2|e^{-\beta\hat H/P}|q_k+\eta_k/2\rangle} \ ,\label{eq:p+}\\
&&\bar p_{k,-}=\frac{\langle q_k-\eta_k/2|\hat p e^{-\beta\hat H/P} |q_{k+1}+\eta_{k+1}/2\rangle }{\langle q_{k}-\eta_k/2|e^{-\beta \hat H/P} |q_k+\eta_k/2\rangle} \ .\label{eq:p-}
\een
No approximation has been made in deriving Eq. (\ref{eq:dc_sst-10+2}).    When evaluated up to only  the leading order of $\beta/P$,  
\ben
&&\bar p_{k,+}\approx \frac{imP}{\hbar\beta}\left (q_{k-1}-q_{k}-\frac{\eta_{k-1}}{2}-\frac{\eta_k}{2} \right )\ ,\label{eq:p+app}\\
&&\bar p_{k,-}\approx \frac{imP}{\hbar \beta}\left (q_{k}-q_{k+1}-\frac{\eta_k}{2}-\frac{\eta_{k+1}}{2}\right) . \label{eq:p-app}
\een
Equation (\ref{eq:dc_sst-10+2}) is in contrast with the expression derived by HA,\cite{hele-jcp138} which amounts to replacing the imaginary time momenta $\tilde p_{k,+}$ and $\tilde p_{k,-}$ with $p_k$ defined by Eq. (\ref{eq:pk_def}).  \rc{ As noted in our earlier work,\cite{jang-jcp144} this is clearly an approximation.} 

In our \rc{earlier} work,\cite{jang-jcp144} we then showed that Eq. (\ref{eq:dc_sst-10+2}) can be evaluated further employing a normal mode transformation used by HA.\cite{hele-jcp138} To this end, it is convenient to introduce\cite{hele-jcp138,jang-jcp144} the following functions:
\ben 
&&T_{k}({\bf q})=\frac{1}{\sqrt{B_P({\bf q})}}\frac{\partial f({\bf q})}{\partial q_k} \ , \label{eq:tk-def}\\
&&B_P({\bf q})=\sum_{k=1}^P \left (\frac{\partial f({\bf q})}{\partial q_k}\right )^2 \ , \label{eq:bnq} \\
&&\tilde \eta_0({\bf q})=\sum_{k=1}^P \eta_k T_{k}({\bf q}) \ .
\een 
Then, employing Eqs. (\ref{eq:p+app})-(\ref{eq:p-app}), and performing integrations over all the modes of $\boldeta$ and ${\bf p}$ perpendicular to the vector formed by $T_k$'s,
we have obtained the following expression:\cite{jang-jcp144}  
\ben
&&k_{HA}Z_a= \frac{P}{2\pi \hbar \beta}\int d{\bf q} \int d \tilde \eta_{_0}({\bf q}) \rho({\bf q},\boldeta_0)\delta (f ({\bf q})-d)\nonumber \\
&&\times \sum_{k=1}^P \frac{\partial f({\bf q})}{\partial q_k} \frac{T_{k-1}({\bf q})+2T_{k}({\bf q})+T_{k+1}({\bf q})}{4}  \ ,\label{eq:dc_sst-10+6}
\een
where $(\boldeta_0)_k=\tilde \eta_0({\bf q})T_k({\bf q})$ and $\rho({\bf q},\boldeta_0)$ can be approximated as
\ben 
\rho({\bf q},\boldeta_0)\approx \rho({\bf q},0) \exp\left \{-\frac{mP}{2\beta\hbar^2}\tilde \eta_0^2({\bf q})-g_{_P}({\bf q})\frac{\tilde \eta_0({\bf q})}{\hbar}\right\} \ . 
\een
In the above expression, $\rho({\bf q},0)$ is the conventional imaginary time path integral representation of the diagonal position element of the canonical density operator \rc{defined as}
\be
\rho({\bf q},0)=\left (\frac{mP}{2\pi\beta \hbar^2}\right)^{P/2} \prod_{k=1}^P e^{-\epsilon V(q_k)-\frac{m}{2\epsilon\hbar^2}(q_k-q_{k+1})^2} \ ,
\ee
with the cyclic boundary condition $q_{_{P+1}}=q_1$, and
\be
g_{_P}({\bf q})=\frac{mP}{2\beta\hbar}\sum_{k=1}^P (q_{k+1}-q_k) T_k({\bf q}) \ . \label{eq:gnq-1}
\ee

In their recent work,\cite{hele-jcp144} HA argued that Eq. (\ref{eq:dc_sst-10+6}) becomes equivalent, in the limit of $P\rightarrow \infty$, to the following version of the RPMD-TST expression:\cite{richardson-jcp131}
 \be
k_{RP}Z_a= \left(\frac{P}{2\pi m\beta}\right)^{1/2}\int d{\bf q}\ \sqrt{B_P({\bf q})} \rho({\bf q},0)\delta (f ({\bf q})-d) \ , \label{eq:rp-tst}
\ee
where $B_P({\bf q})$ is the normalization factor defined by Eq. (\ref{eq:bnq}) and, through Eq. (\ref{eq:tk-def}), can also be expressed as 
\be
\sqrt{B_P({\bf q})}= \sum_{k=1}^P \frac{\partial f({\bf q})}{\partial q_k} T_{k}({\bf q})  \ .\label{eq:bp-rel}
\ee

In order for Eqs. (\ref{eq:dc_sst-10+6}) and (\ref{eq:rp-tst}) to be equivalent in the limit of $P\rightarrow \infty$, as claimed by HA,\cite{hele-jcp144} contributions of $T_{k+1}({\bf q})-T_k({\bf q})$ and $g_{_P}({\bf q})$ in the former should become negligible compared to others in that limit.  To be more precise, the two become equivalent if the following conditions are satisfied:
\ben
&&\lim_{P\rightarrow \infty} (T_{k+1}({\bf q})-T_k({\bf q}))\sqrt{P} =0 \ ,\label{eq:equiv-1}\\
&&\lim_{P\rightarrow \infty} g_{_P}({\bf q})/\sqrt{P}=0 \ .\label{eq:equiv-2}
\een
For the type of smooth dividing surfaces assumed by HA,\cite{hele-jcp138,hele-jcp144} the above conditions are satisfied. However, in the following section, we show that the analysis does not hold true for more general cyclically invariant dividing surfaces in the imaginary time path integral space. 

\section{Dependence of HA-QTST on Dividing Surface}
Because of the cyclic invariance of the summation in Eq. (\ref{eq:gnq-1}), $g_{_P}({\bf q})$ can also be expressed as follows.
\ben
&&g_{_P}({\bf q})=\frac{mP}{2\beta \hbar} \sum_{k=1}^P q_k \left ( T_{k-1}({\bf q}) -T_k({\bf q})\right)  \nonumber \\
&&\hspace{.2in}=\frac{mP}{2\beta\hbar} \frac{1}{\sqrt{B_P({\bf q})}} \sum_{k=1}^P q_k \left (\frac{\partial f({\bf q})}{\partial q_{k-1}}-\frac{\partial f ({\bf q})}{\partial q_k}\right) \ ,\label{eq:gnq-2}
\een
\rc{where Eq. (\ref{eq:tk-def}) has been used in the second equality.}
The main argument made by HA\cite{hele-jcp144} is that  $T_{k+1}({\bf q})-T_k({\bf q}) =O(P^{-1})$ and that $\lim_{P\rightarrow \infty} g_{_P}({\bf q})=0$, for which Eqs. (\ref{eq:equiv-1}) and (\ref{eq:equiv-2}) are satisfied and  Eq. (\ref{eq:dc_sst-10+6}) reduces to Eq. (\ref{eq:rp-tst}) after Gaussian integration over $\tilde\eta_0({\bf q})$.   However, such an agreement is true only for a certain simple class of dividing surfaces in which the contribution of the quantum kinetic energy is vanishingly small, and hence so are the quantum path fluctuations.  For general cyclically invariant dividing surfaces, this is not true.  This can be shown by demonstrating results for two specific examples of the dividing surface as described below.

\subsection{Dividing surface containing a quadratic form in the Fourier modes of the imaginary time path}
\rc{Generalizing the dividing surfaces introduced by Althorpe and coworkers,\cite{richardson-jcp131,hele-jcp138} we here consider a class of cyclically invariant dividing surfaces defined as}
\begin{eqnarray}
f({\bf q})&=&\frac{\cos\phi}{P}\sum_{j=1}^P q_j+\frac{\sqrt{2}\sin \phi}{P} L_n({\bf q}) \ , \label{eq:fq-1}
\end{eqnarray}
where $\phi$ \rc{is a real phase factor that can be} determined or optimized separately, $n$ is a nonnegative integer less than or equal to $P$, and 
\begin{eqnarray}
L_n({\bf q})=  \left (\sum_{j=1}^P \sum_{j'=1}^Pe^{i2\pi n (j-j')/P} q_jq_{j'} \right)^{1/2} \ .
\end{eqnarray}
The above function corresponds to the norm of the $n$th Fourier mode of the cyclic imaginary time path.     \rc{It is worthwhile to note here that there can be additional restriction for $\phi$ in order for $f({\bf q})$ to be a genuine dividing surface in the path integral space.  An obvious condition is that $\cos \phi$ cannot be zero or close to it because the information on the average position of ring polymers gets lost in such case.  With this issue clarified and under the assumption that a physically appropriate choice of $\phi$ can be made, below we examine whether the conditions of Eqs. (\ref{eq:equiv-1}) and (\ref{eq:equiv-2}) are indeed satisfied for any choice of $n$. }

\rc{First, taking derivative of Eq. (\ref{eq:fq-1}) with respect to $q_k$, we find that} 
\begin{eqnarray} 
\frac{\partial f ({\bf q})}{\partial q_k} =\frac{\cos \phi}{P}+\frac{\sqrt{2}\sin \phi}{P L_n({\bf q})}\sum_{j=1}^N \cos \left (\frac{2\pi n(k-j)}{P}\right) q_j \ . \label{eq:partial_f}
\end{eqnarray}
\rc{Using the above expression in Eq. (\ref{eq:bnq}), we also fine that}  
\be 
B_P({\bf q})=\frac{1}{P} \ .
\ee
Therefore, based on the definitions of Eqs. (\ref{eq:tk-def}) and (\ref{eq:gnq-1}), and after employing standard trigonometric manipulation, \rc{we obtain the following general expressions:}
\ben
&&T_{k-1}({\bf q})-T_{k} ({\bf q})\nonumber \\
&&\hspace{.2in}=\sqrt{\frac{2}{P}}\frac{\sin \phi}{L_n({\bf q})}\left \{ \sin (\frac{2\pi n}{P})\sum_{j=1}^P \sin \left (\frac{2\pi n (k-j)}{P}\right) q_j\right . \nonumber \\
&&\hspace{.4in} \left .-2\sin^2(\frac{\pi n}{P})\sum_{j=1}^P \cos \left (\frac{2\pi n (k-j)}{P}\right) q_j\right\} \ , \label{eq:dtk_q}\\
&&g_{_P}({\bf q})=-\frac{m\sin \phi}{\beta\hbar} \sqrt{2P}\sin^2(\frac{\pi n}{P})L_n({\bf q})\ . \label{eq:gnq-new}
\een
\rc{Employing} the fact that $L_n({\bf q})=O(P)$, \rc{from Eq. (\ref{eq:dtk_q})}, we identify the following three types of scaling behavior depending on the value of $n$:  
\begin{equation}
T_{k-1}({\bf q}) -T_k ({\bf q})=\left \{ \begin{array}{lll} O(P^{-1/2})&,&  \mbox{for }n =O(P) \\
                             O(P^{-1})&,& \mbox{for }n=O(P^{1/2}) \\
                            O(P^{-3/2})&, & \mbox{for }n =O(1) 
                            \end{array} \right .     \label{eq:td-order}
\end{equation}
Similarly, \rc{from Eq. (\ref{eq:gnq-new})}, \rc{we find} the following three types of scaling behavior: 
\begin{equation}
g_{_P}({\bf q})=\left \{ \begin{array}{lll} O(P^{3/2})&,&  \mbox{for }n =O(P) \\
                            O(P^{1/2})&, &\mbox{for }n=O(P^{1/2}) \\
                            O(P^{-1/2})&, & \mbox{for }n =O(1) 
                            \end{array} \right .    \label{eq:gn-order}
\end{equation}

The results shown in Eqs. (\ref{eq:td-order}) and (\ref{eq:gn-order}) confirm that the conclusion\cite{hele-jcp144} by HA is true only for $n=O(1)$.  In fact, this limitation was noted briefly in the Appendix A of Ref. \onlinecite{hele-jcp144}, but has not been stated anywhere else.  For the case of $n=O(P^{1/2})$, the term involving $g_{_P}({\bf q})$ does not vanish but remains finite.  Therefore, the difference between HA-QTST and RPMD-TST can be accounted for by a finite factor. For $n=O(P)$,  both the terms involving $T_{k-1}({\bf q})-T_{k}({\bf q})$ and $g_{_P}({\bf q})$ cause HA-QTST to be different from RPMD-TST.  In fact, in this case, it is not clear whether HA-QTST results in a finite value.   Neglecting these cases amount to projecting the imaginary time path integral space onto the subspace where the quantum kinetic energy is not fully accounted for, which it seems cannot be overlooked in any QTST that aspires to more fully incorporate quantum effects.

\rc{As an example, let us consider the case where $P$ is even and $n=P/2$.  Equation (\ref{eq:partial_f}) in this case becomes
\ben
\frac{\partial f({\bf q})}{\partial q_k}=\frac{\cos \phi}{P}-(-1)^k\frac{\sqrt{2}\sin \phi}{2P} \frac{(Q_1-Q_2)}{\sqrt{(Q_1-Q_2)^2}}\ , 
\een
where $Q_1=\sum_{j=1}^{P/2} q_{2j-1}$ and $Q_2=\sum_{j=1}^{P/2}q_{2j}$.
Using the above expression in Eq. (\ref{eq:bnq}), we also find that 
\ben 
B_P({\bf q})=\sum_{k=1}^P \left (\frac{\partial f}{\partial q_k}\right)^2=\frac{1}{P}\left (\cos^2 \phi+\frac{1}{2}\sin^2\phi\right) \ .
\een
Using the above expression in Eq. (\ref{eq:tk-def}), we can calculate that 
\begin{eqnarray}
|T_{k+1}({\bf q})-T_k({\bf q})|=\frac{1}{\sqrt{2P}} \frac{|\sin \phi|}{(\cos^2\phi+\sin^2 \phi/2)^{1/2}} \ .  \label{eq:dtq-ex1}
\end{eqnarray}
In addition, from Eq. (\ref{eq:gnq-2}), we also find that 
\ben
g_{_P}({\bf q})=-\frac{m}{2\beta\hbar} \frac{\sqrt{P}\sin \phi}{(\cos^2\phi+\sin^2\phi/2)}|Q_2-Q_1| = O(P^{3/2}) \ . \label{eq:gp-ex1}
\een
The above two expressions, Eqs. (\ref{eq:dtq-ex1}) and (\ref{eq:gp-ex1}), serve as explicit demonstration of the scaling behavior of Eqs. (\ref{eq:td-order}) and (\ref{eq:gn-order}) for $n=P/2$ and thus prove that both conditions of Eqs. (\ref{eq:equiv-1}) and (\ref{eq:equiv-2}) are indeed violated.  
 } 

\begin{figure}
\includegraphics[width=3.2in]{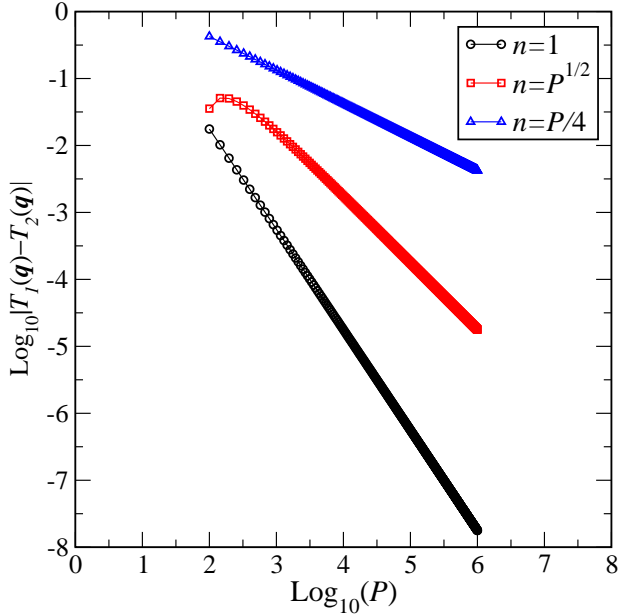}
\caption{Dependences of the absolute value of Eq. (\ref{eq:dtk-ex1}) with $k=2$ and $\alpha=0$ on the number of discretization $P$ in Log-Log scale for three specific examples of $n=1$, $P^{1/2}$, and $P/4$.}
\end{figure}

\rc{As an alternative and more general example, let us consider a specific realization of the cyclic path ${\bf q}$ along which $f({\bf q})$ remains constant as follows:} 
\be
q_j=q_0+\sqrt{2}A_n \sin( \frac{2\pi n j}{P}+\alpha)\ ,
\ee
\rc{where $q_0$, $A_n$, and $\alpha$ are fixed parameters.}  For this choice, it can be shown that
\ben
&&f({\bf q})=q_0\cos\phi +A_n\sin \phi \ , \label{eq:fbq-ex1}\\
&&L_n({\bf q})=\frac{A_nP}{\sqrt{2}}  \ , \label{eq:ln-ex1}\\
&&\frac{\partial f({\bf q})}{\partial q_k}=\frac{\cos \phi}{P}+\frac{\sqrt{2}\sin \phi}{P} \sin (\frac{2\pi n k}{P}+\alpha) \ . \label{eq:par-fq-ex1}
\een
Inserting Eq. (\ref{eq:ln-ex1}) into Eq. (\ref{eq:gnq-new}), we find that
\be
g_{_P}({\bf q})=-\frac{m\sin \phi}{\beta\hbar} A_n \sin^2 (\frac{\pi n}{P}) P^{3/2} \ ,
\ee
which clearly follows the scaling behavior of Eq. (\ref{eq:gn-order}).  Equation (\ref{eq:dtk_q}) \rc{for this cyclic path} can also be calculated and simplifies to
\ben 
&&T_{k-1}({\bf q})-T_k({\bf q})=\frac{2\sqrt{2}}{\sqrt{P}}\left \{ \sin (\frac{2\pi n}{P})\cos(\frac{2\pi n k}{P} +\alpha)\right .\nonumber \\
&&\hspace{1 in}\left .-\sin^2(\frac{\pi n}{P}) \sin(\frac{2\pi n k}{P}+\alpha)\right\} \ . \label{eq:dtk-ex1}
\een
The above expression follows the scaling behavior shown in Eq. (\ref{eq:td-order}).  Figure 1  shows this behavior more clearly for $k=2$ and for three specific choices of $n=1$, $\sqrt{P}$, and $P/4$ assuming $\alpha=0$.

Ultimately, what contributes to the difference between HA-QTST and RPMD-TST is the difference of the summation in Eq. (\ref{eq:dc_sst-10+6}) and that in Eq. (\ref{eq:rp-tst}).    This can also be calculated explicitly as follows:
\ben
&&\frac{1}{4}\sum_{k=1}^P \frac{\partial f ({\bf q})}{\partial q_k}\left \{ (T_{k-1}({\bf q})-T_k({\bf q}))+(T_{k+1}({\bf q})-T_k({\bf q}))\right\} \nonumber \\
&&\hspace{.5in}=\frac{1}{\sqrt{P}} \sin^{2}(\frac{\pi n}{P})\left \{ 3\cos^2(\frac{\pi n}{P})-1\right\} \ .
\een
The above expression shows that the actual difference between the two sums decreases further for $n=O(1)$ and $n=O(P^{1/2})$ due to some cancellation, but the scaling behavior predicted from Eq. (\ref{eq:td-order}) remains the same for $n=O(P)$.

\subsection{Dividing surface containing a quadratic form defined in terms of the imaginary time path coordinates}
We here consider another kind of cyclically invariant dividing surface defined as
\begin{eqnarray}
f({\bf q})&=&\frac{\cos\phi}{P}\sum_{j=1}^P q_j+\frac{\sin \phi}{R(n)} D_n({\bf q}) \ , \label{eq:fq-2}
\end{eqnarray}
where 
\be
D_n({\bf q})=\left (\sum_{j=1}^P(q_j-q_{j+n})^2\right)^{1/2} \ . 
\ee
Because of the cyclic boundary condition for $q_j$, this function is invariant with respect to cyclic permutation.     In Eq. (\ref{eq:fq-2}), $R(n)$ is an appropriate normalization factor that makes $D_n({\bf q})/R(n)$ an order of unity.  For example, for $n=1$, $R(1)=O(1)$.  For $n\sim P/2$, $R(n)=O(\sqrt{P})$.  

For the dividing surface defined above, 
\be
\frac{\partial f({\bf q})}{\partial q_k}=\frac{\cos\phi}{P}+\frac{\sin \phi}{R(n)D_n({\bf q})} (2q_k-q_{k+n}-q_{k-n}) \ ,
\ee
and it can be shown that
\ben 
&&B_P({\bf q})=\frac{\cos^2\phi}{P}\nonumber \\
&&+\frac{\sin^2\phi}{R(n)^2D_n({\bf q})^2}\sum_{k=1}^P(2q_k-q_{k+n}-q_{k-n})^2 \ , \\
&&T_{k-1}({\bf q})-T_k({\bf q})=\frac{\sin \phi}{R(n)\sqrt{B_P({\bf q})}D_n({\bf q})}\Big \{ 2(q_{k-1}-q_k)\nonumber \\
&&\hspace{.4in} -(q_{k-1+n}-q_{k+n})-(q_{k-1-n}-q_{k-n})\Big \} \ ,\\
&&g_{_P}({\bf q})=\frac{mP\sin \phi}{2\beta\hbar R(n)\sqrt{B_P({\bf q})}D_n({\bf q})}\sum_{k=1}^P q_k \Big \{ 2(q_{k-1}-q_k)\nonumber \\
&&\hspace{.4in} -(q_{k-1+n}-q_{k+n})-(q_{k-1-n}-q_{k-n})\Big \} \ .
\een

For the case where $n=1$, $D_1({q}) =O(1)$ and $R(1)=O(1)$.  Therefore,  assuming that $q_{j-1}-q_j=O(P^{-1/2})$ for $j=k$, $k-n$, and $k+n$, we obtain the following estimates. 
\ben 
&&B_P({\bf q})=O(1)\ , \\ 
&&T_{k-1}({\bf q})-T_k({\bf q})=O(P^{-1/2})\ , \\
&&g_{P}({\bf q})=O(P^{3/2})
\een
This is similar to the case of $n=O(P)$ for the dividing surface in the previous subsection, for which HA-QTST may not be well defined.  

On the other hand, for $n\sim P/2$, $D_n({\bf q}) =O(P^{1/2})$ and $R(n)=O(P^{1/2})$.   As a result, 
\ben 
&&B_P({\bf q})=O(P^{-1}) \ ,\\
&&T_{k-1}({\bf q})-T_k({\bf q})=O(P^{-1}) \ , \\
&&g_P({\bf q})=O(P) \ .
\een
The behavior shown above is a mixture of those for $n=O(P)$ and $n=O(P^{1/2})$ for the dividing surface in the previous subsection, and also renders HA-QTST ill-defined.     

The analysis in this subsection thus demonstrates that, for the case where the dividing surface function mixes different Fourier modes of the imaginary time path in a nonlinear manner, the assumption of the analysis by HA\cite{hele-jcp144} does not hold true in general.  As a result, our exact evaluation\cite{jang-jcp144} of HA-QTST\cite{hele-jcp138} expression becomes different from the RPMD-TST expression.\cite{richardson-jcp131}

\section{Concluding Remarks}  
In this work, we have presented a new and detailed analysis of  the effect of general cyclic imaginary time path integral dividing surfaces on the corresponding result of HA-QTST.\cite{hele-jcp138}   What is believed to make HA-QTST unique is the fact that it is defined for an arbitrary form of   cyclically invariant dividing surface constructed in the space of the imaginary time path integral.  However, HA-QTST in its present form does not offer any independent and self-consistent prescription of  the best choice out of the infinite number of dividing surface possibilities available, and its implementation practice has been confined to the dividing surfaces of the type in Eq. (\ref{eq:fq-1}) with $n=1$ followed by a limited variational optimization.   This is different from  the instanton theory,\cite{affleck-prl46,gillan-jp-c20} which \rc{provides a definition} of the dividing surface as the saddle point within the semiclassical approximation for the imaginary time action and can also be \rc{connected to} the PI-QTST.\cite{cao-jcp105,jang-jpca103-2}  Although HA-QTST has been shown to work well for the kind of dividing surface given by Eq. (\ref{eq:fq-1}) with $n=1$, and it becomes equivalent to RPMD-TST for $n=O(1)$, the analysis presented in this work clarifies that different outcomes are expected for more general dividing surfaces.  Along with our previous work,\cite{jang-jcp144} which was focused more on the underlying quantum mechanical expressions and approximations inherent in HA-QTST, the present work again emphasizes the major challenges in developing a unique and well defined QTST while utilizing the full quantum free energy expression defined in the imaginary time path integral space.  

Regardless of our analyses presented here and in a recent work,\cite{jang-jcp144} one can \rc{justifiably} argue that HA-QTST\cite{hele-jcp138} or RPMD-TST\cite{craig-jcp123} for a smooth dividing surface defined in the imaginary time path integral space is useful \rc{in  practice.}  However, one should not misinterpret \rc{this degree of usefulness} as a quantum mechanical validation of some key assumptions and approximations behind HA-QTST and RPMD-TST.  

We conclude this work by offering a few \rc{additional} insights into RPMD-TST and HA-QTST. 
First, it is relatively easy to understand RPMD-TST.\cite{craig-jcp123}  As is clear from Eq. (\ref{eq:rp-tst}), RPMD-TST for a general dividing surface \rc{is the} relative thermal weight for the dividing surface multiplied by the average ``classical" flux across it.  Therefore, this can be seen \rc{as an} extension of the PI-QTST above the crossover temperature\cite{voth-jcp91} for a general dividing surface defined in the imaginary time path integral space.  The fact that this limit is obtained from the RPMD rate theory\cite{craig-jcp122,craig-jcp123} is the result of the principles governing classical mechanics and does not \rc{have a quantum} mechanical origin.

\rc{The implications inherent in HA-QTST\cite{hele-jcp138} are more subtle as} its starting point, GKSCF, already \rc{takes} both classical and ensemble concepts and encodes them into the theory.  Unlike ring polymer averages defined in the RPMD rate theory, which can be considered as ensemble averages of actual physical observables, the GKSCF defines a function of imaginary time paths (a dividing surface in the path integral space) as the argument of the population measurement.  Can this kind of population be measured? Such a measurement requires measuring positions for the quantum canonical ensemble and then giving a weight only when a nonlinear (in general) function constructed from the measured positions meets the criterion of the population.   This is not in general an ensemble average of a physical measurement but rather a subset of the ensemble of measured values that are selected and weighted in certain way.  Even if such a measurement were possible, assuming  a time correlation of such a quantity exists implies that it does not disturb the initial quantum state, which is why its time derivative has a finite $t=0_+$ limit.  Thus, GKSCF \rc{is different} from the original exact Kubo-transformed time correlation function\cite{kubo-jpsj-12,kubo-jpsj-12-2} which represents a well-defined response of a quantum ensemble following a physical perturbation.  

The fact that HA-QTST reduces to RPMD-TST for only smooth dividing surfaces, but may \rc{ not be} well defined for general cyclic dividing surfaces in the path integral space as demonstrated in this work, also reveals an additional \rc{ feature} of the theory.  This result means that HA-QTST becomes equivalent to RPMD-TST only when the effect of operator ordering along the dividing surface of the imaginary time path integral becomes vanishingly small.   In practice, confining the \rc{ dividing} surface to smooth functions ensures that it remains close to the instanton trajectory below the crossover temperature, which may be sufficient.   \rc{Interestingly}, HA-QTST, in its practice of using a smooth dividing surface, may be viewed as a heuristic approximation that interpolates the PI-QTST above the crossover temperature\cite{voth-jcp91} and the instanton theory below the crossover temperature,\cite{affleck-prl46} but without a change in the pre-exponential factor \rc{ (a change that can be argued to be ``needed"\cite{cao-jcp105})}.  Alternatively, \rc{ HA-QTST might be considered to be} a semi-analytical approximation for a numerical \rc{ approach developed some years ago} that proposes\cite{mills-cpl278} to calculate the exact saddle point of the action in the full imaginary time path integral space.\cite{cao-jcp105}  

\acknowledgments
SJ acknowledges the support for this research from  the National Science Foundation  (CHE-1362926) and the Office of Basic Energy Sciences, Department of Energy (DE-SC0001393).  GAV acknowledges the support of the National Science Foundation (NSF) through grant CHE-1465248.  \rc{ We thank David Reichman for discussion in the early stage of this work and Eli Pollak for helpful comments and suggesting detailed consideration of the case $n=P/2$.}

\end{document}